\documentclass[review]{elsarticle}

\usepackage{lineno,hyperref}
\usepackage{graphicx,amsmath,amssymb,amsthm,bm}
\modulolinenumbers[5]

\journal{}

\begin{document}

\begin{frontmatter}

\title{Symmetric equilibrium of multi-agent reinforcement learning in repeated prisoner's dilemma}

\author{Yuki Usui}
\address{Faculty of Science, Yamaguchi University, Yamaguchi 753-8511, Japan}
\ead{i007de@yamaguchi-u.ac.jp}

\author{Masahiko Ueda\corref{mycorrespondingauthor}}
\address{Graduate School of Sciences and Technology for Innovation, Yamaguchi University, Yamaguchi 753-8511, Japan}
\ead{m.ueda@yamaguchi-u.ac.jp}

\begin{abstract}
We investigate the repeated prisoner's dilemma game where both players alternately use reinforcement learning to obtain their optimal memory-one strategies.
We theoretically solve the simultaneous Bellman optimality equations of reinforcement learning.
We find that the Win-stay Lose-shift strategy, the Grim strategy, and the strategy which always defects can form symmetric equilibrium of the mutual reinforcement learning process amongst all deterministic memory-one strategies.
\end{abstract}


\end{frontmatter}


\section{Introduction}
\label{sec:introduction}
The prisoner's dilemma game describes a dilemma where rational behavior of each player cannot achieve a favorable situation for both players \cite{RCO1965}.
In the game, each player chooses cooperation or defection.
Each player can obtain more payoff by taking defection than by taking cooperation regardless of the opponent's action.
Then, mutual defection is realized as a result of rational thought of both players, while payoffs of both players increase when both players choose cooperation.
Although the Nash equilibrium of the one-shot game is mutual defection, when the game is infinitely repeated, it has been known that mutual cooperation can be achieved as the Nash equilibrium.
This fact is known as the folk theorem.
Because the repeated version of the prisoner's dilemma game is also simple, it has substantially been investigated \cite{HCN2018}.

Recently, reinforcement learning technique attracts much attentions in the context of game theory \cite{Rap1967,SanCri1996,SAF2002,HuWel2003,GalFar2013,HTM2015,HKJKGC2017,BDK2019,FujKan2019}.
In reinforcement learning, a player gradually learns his/her optimal strategy against his/her opponents.
Both learning by a single player and learning by several players have been investigated.
Because rationality of players is bounded in reality, modeling of players as learning agents is crucial \cite{BisNai2000}.
It is also significant in the context of reinforcement learning, since the original reinforcement learning was formulated for Markov decision process with stationary environments \cite{SutBar2018}.
Because the existence of multiple agents in game theory leads to non-stationarity of environments for each player, the standard application of reinforcement learning to games breaks down \cite{SanCri1996,BBD2008}, and further theoretical understanding of reinforcement learning in game theory is needed.
Moreover, since the acquisition process of optimal strategies in reinforcement learning is generally different from that in evolutionary game theory \cite{SmiPri1973}, accumulating knowledge about equilibrium in each learning dynamics is needed.

In this paper, we investigate the situation where both players alternately learn their optimal strategies by using reinforcement learning in the repeated prisoner's dilemma game.
We theoretically derive equilibrium points of mutual reinforcement learning where both players take the same deterministic strategy.
We find that the strategy which always defects (All-$D$), the Win-stay Lose-Shift (WSLS) strategy \cite{NowSig1993}, and the Grim strategy can form such symmetric equilibrium amongst all memory-one deterministic strategies.

This paper is organized as follows.
In Section \ref{sec:model}, we introduce the repeated prisoner's dilemma game, and players using reinforcement learning.
In Section \ref{sec:results}, we theoretically derive deterministic optimal strategies against the strategy of a learning opponent.
In Section \ref{sec:numerical}, we provide numerical results by using Q-learning which support our theoretical results.
Section \ref{sec:conclusion} is devoted to conclusion.

\section{Model}
\label{sec:model}
We consider the repeated prisoner's dilemma game \cite{Rap1967}.
There are two players in the game, and each player is described as $1$ and $2$.
Each player chooses cooperation ($C$) or defection ($D$) on every trial.
The action of player $a$ is written as $\sigma_a \in \{ C, D \}$, and we collectively write $\bm{\sigma}:=\left( \sigma_1, \sigma_2 \right)$.
The payoff of player $a\in \{ 1, 2 \}$ when the state is $\bm{\sigma}$ is described as $r_a\left( \bm{\sigma} \right)$.
The payoffs in the prisoner's dilemma game are defined as
\begin{eqnarray}
 \left(
\begin{array}{cc}
r_1 \left( C, C \right), r_2 \left( C, C \right) & r_1 \left( C, D \right), r_2 \left( C, D \right) \\
r_1 \left( D, C \right), r_2 \left( D, C \right) & r_1 \left( D, D \right), r_2 \left( D, D \right)
\end{array}
\right) &=& \left(
\begin{array}{cc}
R, R & S, T \\
T, S & P, P
\end{array}
\right)
\end{eqnarray}
with $T>R>P>S$ and $2R>T+S$.
We consider the situation where both players use memory-one strategies.
The memory-one strategy of player $a$ is described as the conditional probability $T_a\left( \sigma_a | \bm{\sigma}^{\prime} \right)$ of taking action $\sigma_a$ when the state in the previous round is $\bm{\sigma}^{\prime}$.
(In this paper, we investigate only the game with perfect monitoring, where players can perfectly observe the actions of both players in the previous round.)
Then, when we define the probability distribution of a state $\bm{\sigma}^{\prime}$ at time $t$ by $P\left( \bm{\sigma}^{\prime}, t \right)$, the time evolution of this system is described as the Markov chain
\begin{eqnarray}
 P\left( \bm{\sigma}, t+1 \right) &=& \sum_{\bm{\sigma}^{\prime}} T\left( \bm{\sigma} | \bm{\sigma}^{\prime} \right) P\left( \bm{\sigma}^{\prime}, t \right)
\end{eqnarray}
with the transition probability
\begin{eqnarray}
 T\left( \bm{\sigma} | \bm{\sigma}^{\prime} \right) &:=& \prod_{a=1}^2 T_a\left( \sigma_a | \bm{\sigma}^{\prime} \right).
\end{eqnarray}
Below we introduce the notation $-a := \{ 1, 2 \} \backslash a$.

We consider the situation that both players learn their strategies by reinforcement learning \cite{SutBar2018}.
We assume that two players alternately learn and update their strategies \cite{Lan2020}, that is, player $1$ first learns her strategy against a fixed initial strategy of player $2$, then player $2$ learns his strategy against the strategy of player $1$, then player $1$ learns her strategy against the strategy of player $2$, and so on.
In other words, the two players infinitely repeat the infinitely repeated game, and their strategies are updated after each repeated game is played and their long-term payoffs are calculated.
We assume that the strategy of player $1$ is updated in $n$-th game with $n=2m-1$ $(m\in \mathbb{N})$ and the strategy of player $2$ is updated in $n$-th game with $n=2m$ $(m\in \mathbb{N})$.
We write the strategies of player $a$ at $n$-th game as $T^{(n)}_a\left( \sigma_a | \bm{\sigma}^{\prime} \right)$.

In reinforcement learning, each player learns mapping (called policy) from a state to his/her action so as to maximize his/her expected future reward.
In our memory-one situation, a state and an action of player $a$ are regarded as the state $\bm{\sigma}^\prime$ in the previous round and the action $\sigma_a$ in the present round, respectively.
We define the action-value function of player $a$ as
\begin{eqnarray}
 Q_a\left( \sigma_a^{(1)}, \bm{\sigma}^{(0)} \right) &:=& \mathbb{E} \left[ \left. \sum_{k=0}^\infty \gamma^k r_a(t+k+1) \right| \sigma_a(t+1)= \sigma_a^{(1)}, \bm{\sigma}(t)=\bm{\sigma}^{(0)} \right], \nonumber \\
 &&
 \label{eq:def_Q}
\end{eqnarray}
where $\gamma$ is a discounting factor satisfying $0\leq \gamma < 1$.
The action $\sigma_a(t)$ represents the action of player $a$ at round $t$.
Similarly, the payoff $r_a(t)$ represents the payoff of player $a$ at round $t$, that is, $r_a(t):= r_a\left( \bm{\sigma}(t) \right)$.
Due to the Markov property, the action-value function $Q$ obeys the Bellman equation against a fixed strategy $T_{-a}$ of the opponent:
\begin{eqnarray}
 Q_a\left( \sigma_a^{(1)}, \bm{\sigma}^{(0)} \right) &=& \sum_{\sigma_{-a}^{(1)}} T_{-a} \left( \sigma_{-a}^{(1)} | \bm{\sigma}^{(0)} \right) r_a \left( \bm{\sigma}^{(1)} \right) \nonumber \\
 && + \gamma \sum_{\sigma_{a}^{(2)}} \sum_{\sigma_{-a}^{(1)}} T_{a} \left( \sigma_{a}^{(2)} | \bm{\sigma}^{(1)} \right) T_{-a} \left( \sigma_{-a}^{(1)} | \bm{\sigma}^{(0)} \right) Q_a \left( \sigma_{a}^{(2)}, \bm{\sigma}^{(1)} \right). \nonumber \\
 &&
\end{eqnarray}
It has been known that the optimal value of $Q$ obeys the following Bellman optimality equation:
\begin{eqnarray}
 Q^*_a\left( \sigma_a^{(1)}, \bm{\sigma}^{(0)} \right) &=& \sum_{\sigma_{-a}^{(1)}} T_{-a} \left( \sigma_{-a}^{(1)} | \bm{\sigma}^{(0)} \right) r_a \left( \bm{\sigma}^{(1)} \right) + \gamma \sum_{\sigma_{-a}^{(1)}} T_{-a} \left( \sigma_{-a}^{(1)} | \bm{\sigma}^{(0)} \right) \max_{\sigma_{a}^{(2)}} Q^*_a \left( \sigma_{a}^{(2)}, \bm{\sigma}^{(1)} \right) \nonumber \\
 &&
 \label{eq:bellman}
\end{eqnarray}
with the support
\begin{eqnarray}
 \mathrm{supp}T_a\left( \left. \cdot \right| \bm{\sigma}^{(0)} \right) &=& \arg \max_{\sigma} Q^*_a \left( \sigma, \bm{\sigma}^{(0)} \right).
 \label{eq:BOE_support}
\end{eqnarray}
In other words, in the optimal policy against $T_{-a}$, player $a$ takes the action $\sigma_a$ which maximizes the value of $Q^*_a\left( \cdot, \bm{\sigma}^{(0)} \right)$ when the state at the previous round is $\bm{\sigma}^{(0)}$.

In sum, in the $(2m-1)$-th game, player $1$ learns $T^{(2m-1)}_1\left( \sigma_1 | \bm{\sigma}^{\prime} \right)$ against $T^{(2m-2)}_2\left( \sigma_2 | \bm{\sigma}^{\prime} \right)$ by calculating $Q^{*(2m-1)}_1 \left( \sigma, \bm{\sigma}^{(0)} \right)$, where $Q^{*(n)}_a \left( \sigma, \bm{\sigma}^{(0)} \right)$ represents the optimal action-value function of player $a$ in the $n$-th game.
In the $2m$-th game, player $2$ learns $T^{(2m)}_2\left( \sigma_2 | \bm{\sigma}^{\prime} \right)$ against $T^{(2m-1)}_1\left( \sigma_1 | \bm{\sigma}^{\prime} \right)$ by calculating $Q^{*(2m)}_2 \left( \sigma, \bm{\sigma}^{(0)} \right)$.
We are interested in the fixed points of the dynamics, that is, $T^{(\infty)}_a\left( \sigma_a | \bm{\sigma}^{\prime} \right)$ and $Q^{*(\infty)}_a \left( \sigma, \bm{\sigma}^{(0)} \right)$.

In this paper, we investigate only situations that the support (\ref{eq:BOE_support}) contains only one action, that is, we investigate only deterministic strategies.
Because the number of deterministic memory-one strategies in the repeated prisoner's dilemma game is sixteen, we check whether each deterministic strategy forms equilibrium or not.

\section{Results}
\label{sec:results}
We consider symmetric solutions of Eq. (\ref{eq:bellman}), that is,
\begin{eqnarray}
 Q^*_1\left( \sigma_1^{(1)}, \bm{\sigma}^{(0)} \right) &=& \sum_{\sigma_{2}^{(1)}} T_{2} \left( \sigma_{2}^{(1)} | \bm{\sigma}^{(0)} \right) r_1 \left( \bm{\sigma}^{(1)} \right) + \gamma \sum_{\sigma_{2}^{(1)}} T_{2} \left( \sigma_{2}^{(1)} | \bm{\sigma}^{(0)} \right) \max_{\sigma_{1}^{(2)}} Q^*_1 \left( \sigma_{1}^{(2)}, \bm{\sigma}^{(1)} \right) \nonumber \\
 &&
 \label{eq:optimal_1}
\end{eqnarray}
with
\begin{eqnarray}
 T_2 \left( C | C, C \right) &=& \mathbb{I} \left( Q^*_1(C, C, C) > Q^*_1(D, C, C) \right) \\
 T_2 \left( C | C, D \right) &=& \mathbb{I} \left( Q^*_1(C, D, C) > Q^*_1(D, D, C) \right) \\
 T_2 \left( C | D, C \right) &=& \mathbb{I} \left( Q^*_1(C, C, D) > Q^*_1(D, C, D) \right) \\
 T_2 \left( C | D, D \right) &=& \mathbb{I} \left( Q^*_1(C, D, D) > Q^*_1(D, D, D) \right)
\end{eqnarray}
where $\mathbb{I}(\cdots)$ is the indicator function that returns $1$ when $\cdots$ holds and $0$ otherwise.
Then, Eq. (\ref{eq:bellman}) becomes
\begin{eqnarray}
 Q^*_1\left( C, C, C \right) &=& \mathbb{I} \left( Q^*_1(C, C, C) > Q^*_1(D, C, C) \right) \left\{ R+\gamma \max_\sigma Q^*_1 (\sigma, C, C) \right\} \nonumber \\
 && \qquad + \mathbb{I} \left( Q^*_1(C, C, C) < Q^*_1(D, C, C) \right) \left\{ S+\gamma \max_\sigma Q^*_1 (\sigma, C, D) \right\} \nonumber \\
 &&
\end{eqnarray}
\begin{eqnarray}
 Q^*_1\left( C, C, D \right) &=& \mathbb{I} \left( Q^*_1(C, D, C) > Q^*_1(D, D, C) \right) \left\{ R+\gamma \max_\sigma Q^*_1 (\sigma, C, C) \right\} \nonumber \\
 && \qquad + \mathbb{I} \left( Q^*_1(C, D, C) < Q^*_1(D, D, C) \right) \left\{ S+\gamma \max_\sigma Q^*_1 (\sigma, C, D) \right\} \nonumber \\
 &&
\end{eqnarray}
\begin{eqnarray}
 Q^*_1\left( C, D, C \right) &=& \mathbb{I} \left( Q^*_1(C, C, D) > Q^*_1(D, C, D) \right) \left\{ R+\gamma \max_\sigma Q^*_1 (\sigma, C, C) \right\} \nonumber \\
 && \qquad + \mathbb{I} \left( Q^*_1(C, C, D) < Q^*_1(D, C, D) \right) \left\{ S+\gamma \max_\sigma Q^*_1 (\sigma, C, D) \right\} \nonumber \\
 &&
\end{eqnarray}
\begin{eqnarray}
 Q^*_1\left( C, D, D \right) &=& \mathbb{I} \left( Q^*_1(C, D, D) > Q^*_1(D, D, D) \right) \left\{ R+\gamma \max_\sigma Q^*_1 (\sigma, C, C) \right\} \nonumber \\
 && \qquad + \mathbb{I} \left( Q^*_1(C, D, D) < Q^*_1(D, D, D) \right) \left\{ S+\gamma \max_\sigma Q^*_1 (\sigma, C, D) \right\} \nonumber \\
 &&
\end{eqnarray}
\begin{eqnarray}
 Q^*_1\left( D, C, C \right) &=& \mathbb{I} \left( Q^*_1(C, C, C) > Q^*_1(D, C, C) \right) \left\{ T+\gamma \max_\sigma Q^*_1 (\sigma, D, C) \right\} \nonumber \\
 && \qquad + \mathbb{I} \left( Q^*_1(C, C, C) < Q^*_1(D, C, C) \right) \left\{ P+\gamma \max_\sigma Q^*_1 (\sigma, D, D) \right\} \nonumber \\
 &&
\end{eqnarray}
\begin{eqnarray}
 Q^*_1\left( D, C, D \right) &=& \mathbb{I} \left( Q^*_1(C, D, C) > Q^*_1(D, D, C) \right) \left\{ T+\gamma \max_\sigma Q^*_1 (\sigma, D, C) \right\} \nonumber \\
 && \qquad + \mathbb{I} \left( Q^*_1(C, D, C) < Q^*_1(D, D, C) \right) \left\{ P+\gamma \max_\sigma Q^*_1 (\sigma, D, D) \right\} \nonumber \\
 &&
\end{eqnarray}
\begin{eqnarray}
 Q^*_1\left( D, D, C \right) &=& \mathbb{I} \left( Q^*_1(C, C, D) > Q^*_1(D, C, D) \right) \left\{ T+\gamma \max_\sigma Q^*_1 (\sigma, D, C) \right\} \nonumber \\
 && \qquad + \mathbb{I} \left( Q^*_1(C, C, D) < Q^*_1(D, C, D) \right) \left\{ P+\gamma \max_\sigma Q^*_1 (\sigma, D, D) \right\} \nonumber \\
 &&
\end{eqnarray}
\begin{eqnarray}
 Q^*_1\left( D, D, D \right) &=& \mathbb{I} \left( Q^*_1(C, D, D) > Q^*_1(D, D, D) \right) \left\{ T+\gamma \max_\sigma Q^*_1 (\sigma, D, C) \right\} \nonumber \\
 && \qquad + \mathbb{I} \left( Q^*_1(C, D, D) < Q^*_1(D, D, D) \right) \left\{ P+\gamma \max_\sigma Q^*_1 (\sigma, D, D) \right\}. \nonumber \\
 &&
\end{eqnarray}
For simplicity, we introduce the following notation:
\begin{eqnarray}
 q_1 &:=& Q^*_1\left( C, C, C \right) \nonumber \\
 q_2 &:=& Q^*_1\left( C, C, D \right) \nonumber \\
 q_3 &:=& Q^*_1\left( C, D, C \right) \nonumber \\
 q_4 &:=& Q^*_1\left( C, D, D \right) \nonumber \\
 q_5 &:=& Q^*_1\left( D, C, C \right) \nonumber \\
 q_6 &:=& Q^*_1\left( D, C, D \right) \nonumber \\
 q_7 &:=& Q^*_1\left( D, D, C \right) \nonumber \\
 q_8 &:=& Q^*_1\left( D, D, D \right).
 \label{eq:def_q}
\end{eqnarray}
We consider the following sixteen situations separately.

\subsection{Case 1: $q_1>q_5$, $q_2>q_6$, $q_3>q_7$, and $q_4>q_8$}
For this case, the strategy obtained by reinforcement learning is the All-$C$ strategy.
The solution of Eq. (\ref{eq:bellman}) is
\begin{eqnarray}
 q_1 = q_2 = q_3 = q_4 &=& \frac{1}{1-\gamma} R \\
 q_5 = q_6 = q_7 = q_8 &=& T + \frac{\gamma}{1-\gamma} R.
\end{eqnarray}
This contradicts with the definition of the game $T>R$.

\subsection{Case 2: $q_1>q_5$, $q_2>q_6$, $q_3>q_7$, and $q_4<q_8$}
The solution of Eq. (\ref{eq:bellman}) is
\begin{eqnarray}
 q_1 = q_2 = q_3 &=& \frac{1}{1-\gamma} R \\
 q_4 &=& S+\frac{\gamma}{1-\gamma}R \\
 q_5 = q_6 = q_7 &=& T + \frac{\gamma}{1-\gamma} R \\
 q_8 &=& \frac{1}{1-\gamma}P.
\end{eqnarray}
This contradicts with the definition of the game $T>R$.

\subsection{Case 3: $q_1>q_5$, $q_2>q_6$, $q_3<q_7$, and $q_4>q_8$}
The solution of Eq. (\ref{eq:bellman}) is
\begin{eqnarray}
 q_1 = q_3 = q_4 &=& \frac{1}{1-\gamma} R \\
 q_2 &=& \frac{1}{1-\gamma}S \\
 q_5 = q_7 = q_8 &=& \frac{1}{1-\gamma} T \\
 q_6 &=& P + \frac{\gamma}{1-\gamma}R.
\end{eqnarray}
This contradicts with the definition of the game $T>R$.

\subsection{Case 4: $q_1>q_5$, $q_2>q_6$, $q_3<q_7$, and $q_4<q_8$}
For this case, the strategy obtained by reinforcement learning is ``Repeat'' \cite{Aki2012}.
The solution of Eq. (\ref{eq:bellman}) is
\begin{eqnarray}
 q_1 = q_3 &=& \frac{1}{1-\gamma} R \\
 q_2 = q_4 &=& \frac{1}{1-\gamma} S \\
 q_5 = q_7 &=& \frac{1}{1-\gamma} T \\
 q_6 = q_8 &=& \frac{1}{1-\gamma} P.
\end{eqnarray}
This contradicts with the definition of the game $T>R$.

\subsection{Case 5: $q_1>q_5$, $q_2<q_6$, $q_3>q_7$, and $q_4>q_8$}
The solution of Eq. (\ref{eq:bellman}) is
\begin{eqnarray}
 q_1 = q_2 = q_4 &=& \frac{1}{1-\gamma} R \\
 q_3 &=& \frac{1}{1-\gamma^2} S + \frac{\gamma}{1-\gamma^2} T \\
 q_5 = q_6 = q_8 &=& \frac{1}{1-\gamma^2} T + \frac{\gamma}{1-\gamma^2} S \\
 q_7 &=& P + \frac{\gamma}{1-\gamma} R.
\end{eqnarray}
This contradicts with $2R>T+S$.

\subsection{Case 6: $q_1>q_5$, $q_2<q_6$, $q_3>q_7$, and $q_4<q_8$}
For this case, the strategy obtained by reinforcement learning is Tit-for-Tat (TFT) \cite{RCO1965,AxeHam1981}.
The solution of Eq. (\ref{eq:bellman}) is
\begin{eqnarray}
 q_1 = q_2 &=& \frac{1}{1-\gamma} R \\
 q_3 = q_4 &=& \frac{1}{1-\gamma^2} S + \frac{\gamma}{1-\gamma^2} T \\
 q_5 = q_6 &=& \frac{1}{1-\gamma^2} T + \frac{\gamma}{1-\gamma^2} S \\
 q_7 = q_8 &=& \frac{1}{1-\gamma} P.
\end{eqnarray}
This solution becomes consistent with the condition of the case only when $T+S=R+P$ and $\gamma=\frac{T-R}{R-S}$.

\subsection{Case 7: $q_1>q_5$, $q_2<q_6$, $q_3<q_7$, and $q_4>q_8$}
For this case, the strategy obtained by reinforcement learning is Win-stay-Lose-Shift (WSLS) \cite{NowSig1993}.
The solution of Eq. (\ref{eq:bellman}) is
\begin{eqnarray}
 q_1 = q_4 &=& \frac{1}{1-\gamma} R \\
 q_2 = q_3 &=& S + \gamma P + \frac{\gamma^2}{1-\gamma} R \\
 q_5 = q_8 &=& T + \gamma P + \frac{\gamma^2}{1-\gamma} R \\
 q_6 = q_7 &=& P + \frac{\gamma}{1-\gamma} R.
\end{eqnarray}
This solution becomes consistent with the condition of the case when $T+P<2R$ and $\gamma>\frac{T-R}{R-P}$.

\subsection{Case 8: $q_1>q_5$, $q_2<q_6$, $q_3<q_7$, and $q_4<q_8$}
For this case, the strategy obtained by reinforcement learning is the Grim strategy.
The solution of Eq. (\ref{eq:bellman}) is
\begin{eqnarray}
 q_1 &=& \frac{1}{1-\gamma} R \\
 q_2 = q_3 = q_4 &=& S + \frac{\gamma}{1-\gamma} P \\
 q_5 &=& T + \frac{\gamma}{1-\gamma} P \\
 q_6 = q_7 = q_8 &=& \frac{1}{1-\gamma} P.
\end{eqnarray}
This solution becomes consistent with the condition of the case when $\gamma>\frac{T-R}{T-P}$.

\subsection{Case 9: $q_1<q_5$, $q_2>q_6$, $q_3>q_7$, and $q_4>q_8$}
For this case, the strategy obtained by reinforcement learning is the anti-Grim strategy.
The solution of Eq. (\ref{eq:bellman}) is
\begin{eqnarray}
 q_1 &=& S + \frac{\gamma}{1-\gamma^2} R + \frac{\gamma^2}{1-\gamma^2} P \\
 q_2 = q_3 = q_4 &=& \frac{1}{1-\gamma^2} R + \frac{\gamma}{1-\gamma^2} P \\
 q_5 &=& \frac{1}{1-\gamma^2} P + \frac{\gamma}{1-\gamma^2} R \\
 q_6 = q_7 = q_8 &=& T + \frac{\gamma}{1-\gamma^2} R + \frac{\gamma^2}{1-\gamma^2} P.
\end{eqnarray}
This contradicts with $\gamma\geq 0$.

\subsection{Case 10: $q_1<q_5$, $q_2>q_6$, $q_3>q_7$, and $q_4<q_8$}
For this case, the strategy obtained by reinforcement learning is anti-Win-stay-Lose-Shift (AWSLS).
The solution of Eq. (\ref{eq:bellman}) is
\begin{eqnarray}
 q_1 = q_4 &=& S + \gamma R + \frac{\gamma^2}{1-\gamma} P \\
 q_2 = q_3 &=& R + \frac{\gamma}{1-\gamma} P \\
 q_5 = q_8 &=& \frac{1}{1-\gamma} P \\
 q_6 = q_7 &=& T + \gamma R + \frac{\gamma^2}{1-\gamma} P.
\end{eqnarray}
This contradicts with $\gamma\geq 0$.

\subsection{Case 11: $q_1<q_5$, $q_2>q_6$, $q_3<q_7$, and $q_4>q_8$}
For this case, the strategy obtained by reinforcement learning is anti-Tit-for-Tat (ATFT).
The solution of Eq. (\ref{eq:bellman}) is
\begin{eqnarray}
 q_1 = q_2 &=& \frac{1}{1-\gamma} S \\
 q_3 = q_4 &=& \frac{1}{1-\gamma^2} R + \frac{\gamma}{1-\gamma^2} P \\
 q_5 = q_6 &=& \frac{1}{1-\gamma^2} P + \frac{\gamma}{1-\gamma^2} R \\
 q_7 = q_8 &=& \frac{1}{1-\gamma} T.
\end{eqnarray}
This contradicts with $\gamma\geq 0$.

\subsection{Case 12: $q_1<q_5$, $q_2>q_6$, $q_3<q_7$, and $q_4<q_8$}
The solution of Eq. (\ref{eq:bellman}) is
\begin{eqnarray}
 q_1 = q_2 = q_4 &=& \frac{1}{1-\gamma} S \\
 q_3 &=& R + \frac{\gamma}{1-\gamma} P \\
 q_5 = q_6 = q_8 &=& \frac{1}{1-\gamma} P \\
 q_7 &=& \frac{1}{1-\gamma} T.
\end{eqnarray}
This contradicts with the definition of the game $P>S$.

\subsection{Case 13: $q_1<q_5$, $q_2<q_6$, $q_3>q_7$, and $q_4>q_8$}
For this case, the strategy obtained by reinforcement learning is anti-Repeat.
The solution of Eq. (\ref{eq:bellman}) is
\begin{eqnarray}
 q_1 = q_3 &=& \frac{1}{1-\gamma^2} S + \frac{\gamma}{1-\gamma^2} T \\
 q_2 = q_4 &=& \frac{1}{1-\gamma^2} R + \frac{\gamma}{1-\gamma^2} P \\
 q_5 = q_7 &=& \frac{1}{1-\gamma^2} P + \frac{\gamma}{1-\gamma^2} R \\
 q_6 = q_8 &=& \frac{1}{1-\gamma^2} T + \frac{\gamma}{1-\gamma^2} S.
\end{eqnarray}
This solution becomes consistent with the condition of the case only when $T+S=R+P$ and $\gamma=1$.

\subsection{Case 14: $q_1<q_5$, $q_2<q_6$, $q_3>q_7$, and $q_4<q_8$}
The solution of Eq. (\ref{eq:bellman}) is
\begin{eqnarray}
 q_1 = q_3 = q_4 &=& \frac{1}{1-\gamma^2} S + \frac{\gamma}{1-\gamma^2} T \\
 q_2 &=& R + \frac{\gamma}{1-\gamma}P \\
 q_5 = q_7 = q_8 &=& \frac{1}{1-\gamma} P \\
 q_6 &=& \frac{1}{1-\gamma^2} T + \frac{\gamma}{1-\gamma^2} S.
\end{eqnarray}
This solution becomes consistent with the condition of the case only when $T+S>2P$ and $\gamma=\frac{P-S}{T-S}$.

\subsection{Case 15: $q_1<q_5$, $q_2<q_6$, $q_3<q_7$, and $q_4>q_8$}
The solution of Eq. (\ref{eq:bellman}) is
\begin{eqnarray}
 q_1 = q_2 = q_3 &=& S + \frac{\gamma}{1-\gamma^2} P + \frac{\gamma^2}{1-\gamma^2} R \\
 q_4 &=& \frac{1}{1-\gamma^2}R + \frac{\gamma}{1-\gamma^2}P \\
 q_5 = q_6 = q_7 &=& \frac{1}{1-\gamma^2}P + \frac{\gamma}{1-\gamma^2}R \\
 q_8 &=& T + \frac{\gamma}{1-\gamma^2} P + \frac{\gamma^2}{1-\gamma^2} R.
\end{eqnarray}
This contradicts with the definition of the game $T>R$.

\subsection{Case 16: $q_1<q_5$, $q_2<q_6$, $q_3<q_7$, and $q_4<q_8$}
For this case, the strategy obtained by reinforcement learning is the All-$D$ strategy.
The solution of Eq. (\ref{eq:bellman}) is
\begin{eqnarray}
 q_1 = q_2 = q_3 = q_4 &=& S + \frac{\gamma}{1-\gamma} P \\
 q_5 = q_6 = q_7 = q_8 &=& \frac{1}{1-\gamma} P.
\end{eqnarray}
This solution is always consistent with the condition of the case.

\subsection{Summary}
From the above subsections, we find that the symmetric solution of the Bellman optimality equation exists in finite regions of the parameter $\gamma$ only for the case 7, 8, and 16.
In other words, only WSLS, the Grim strategy, and the All-$D$ strategy can form the symmetric equilibrium of mutual reinforcement learning.
TFT does not form symmetric equilibrium.
(The optimal strategy against TFT is investigated in detail in \ref{app:optimal}.)
The results are summarized in Table \ref{table:summary}, where the strategy vector of player $1$ is defined by
\begin{eqnarray}
 \bm{T}_1 (C) &:=&  \left(
\begin{array}{c}
T_1 \left( C | C, C \right) \\
T_1 \left( C | C, D \right) \\
T_1 \left( C | D, C \right) \\
T_1 \left( C | D, D \right)
\end{array}
\right).
\end{eqnarray}
\begin{table}[tb]
\scalebox{0.8}{
  \begin{tabular}{|c|c|c|c|c|c|c|c|} \hline
   number & $q_1\lessgtr q_5$ & $q_2\lessgtr q_6$ & $q_3\lessgtr q_7$ & $q_4\lessgtr q_8$ & strategy $\bm{T}_1 (C)$ & name & Equilibrium? \\ \hline
   Case 1 & $>$ & $>$ & $>$ & $>$ & $(1, 1, 1, 1)^\mathsf{T}$ & All-$C$ & No \\
   Case 2 & $>$ & $>$ & $>$ & $<$ & $(1, 1, 1, 0)^\mathsf{T}$ & & No \\
   Case 3 & $>$ & $>$ & $<$ & $>$ & $(1, 1, 0, 1)^\mathsf{T}$ & & No \\
   Case 4 & $>$ & $>$ & $<$ & $<$ & $(1, 1, 0, 0)^\mathsf{T}$ & Repeat & No \\
   Case 5 & $>$ & $<$ & $>$ & $>$ & $(1, 0, 1, 1)^\mathsf{T}$ & & No \\
   Case 6 & $>$ & $<$ & $>$ & $<$ & $(1, 0, 1, 0)^\mathsf{T}$ & TFT & No in general \\
   Case 7 & $>$ & $<$ & $<$ & $>$ & $(1, 0, 0, 1)^\mathsf{T}$ & WSLS & Yes for $\gamma>\frac{T-R}{R-P}$ \\
   Case 8 & $>$ & $<$ & $<$ & $<$ & $(1, 0, 0, 0)^\mathsf{T}$ & Grim & Yes for $\gamma>\frac{T-R}{T-P}$ \\
   Case 9 & $<$ & $>$ & $>$ & $>$ & $(0, 1, 1, 1)^\mathsf{T}$ & anti-Grim & No \\
   Case 10 & $<$ & $>$ & $>$ & $<$ & $(0, 1, 1, 0)^\mathsf{T}$ & AWSLS & No \\
   Case 11 & $<$ & $>$ & $<$ & $>$ & $(0, 1, 0, 1)^\mathsf{T}$ & ATFT & No \\
   Case 12 & $<$ & $>$ & $<$ & $<$ & $(0, 1, 0, 0)^\mathsf{T}$ & & No \\
   Case 13 & $<$ & $<$ & $>$ & $>$ & $(0, 0, 1, 1)^\mathsf{T}$ & anti-Repeat & No in general \\
   Case 14 & $<$ & $<$ & $>$ & $<$ & $(0, 0, 1, 0)^\mathsf{T}$ & & No in general \\
   Case 15 & $<$ & $<$ & $<$ & $>$ & $(0, 0, 0, 1)^\mathsf{T}$ & & No \\
   Case 16 & $<$ & $<$ & $<$ & $<$ & $(0, 0, 0, 0)^\mathsf{T}$ & All-$D$ & Yes \\ \hline
  \end{tabular}
  }
  \caption{Summary of the results.}
  \label{table:summary}
\end{table}

The reason why TFT does not form equilibrium can be intuitively explained as follows.
We consider the situation that player $2$ uses TFT, and the previous state was $(D, C)$.
By definition, the action-value function (\ref{eq:def_Q}) represents expected future reward when the previous state was $\bm{\sigma}^{(0)}$.
If the strategy of player $1$ is also TFT, the sequence
\begin{eqnarray}
 (D,C) \rightarrow (C,D) \rightarrow (D,C) \rightarrow (C,D) \rightarrow \cdots
\end{eqnarray}
is realized.
If the strategy of player $1$ is All-$C$, the sequence
\begin{eqnarray}
 (D,C) \rightarrow (C,D) \rightarrow (C,C) \rightarrow (C,C) \rightarrow \cdots
\end{eqnarray}
is realized.
Because of $T+S<2R$, the latter results in larger total payoff than the former.
Therefore, as explained in \ref{app:optimal}, the optimal strategy against TFT is not TFT.

One may recall that All-$D$, WSLS and Grim form subgame perfect equilibria in the repeated prisoner's dilemma game, but TFT dose not \cite{IFN2007}.
Therefore, our reinforcement learning equilibrium seems to be similar to subgame perfect equilibrium.
In fact, the above discussion that TFT does not form reinforcement learning equilibrium is similar to the discussion that TFT does not form subgame perfect equilibrium.
However, we expect that reinforcement learning equilibrium is weaker than subgame perfect equilibrium, since, in the definition of subgame perfect equilibrium, arbitrary histories are considered.
Relation between them should be clarified in future.

\section{Numerical results}
\label{sec:numerical}
In this section, we check the theoretical results in the previous section by numerical simulation.
We use Q-learning \cite{SutBar2018} as a method of reinforcement learning.
In Q-learning, the optimal action-value function of the agent $a$ against a fixed strategy of the agent $-a$ is learned through the following update rule:
\begin{eqnarray}
 Q^{(t+1)}_a\left( \sigma_a^{(1)}, \bm{\sigma}^{(0)} \right) &=& Q^{(t)}_a\left( \sigma_a^{(1)}, \bm{\sigma}^{(0)} \right) + \eta \left( r_a + \gamma \max_{\sigma_a^{(2)}}Q^{(t)}_a\left( \sigma_a^{(2)}, \bm{\sigma}^{(1)} \right) - Q^{(t)}_a\left( \sigma_a^{(1)}, \bm{\sigma}^{(0)} \right) \right), \nonumber \\
 &&
\end{eqnarray}
where $r_a$ is the reward by taking action $\sigma_a^{(1)}$ when the state is $\bm{\sigma}^{(0)}$, and $\bm{\sigma}^{(1)}$ is the next state.
The parameter $\eta$ is called the learning rate.
Here, we assume that, in each step, the agent $a$ chooses the action $\sigma_a^{(1)}$ by using $\epsilon$-greedy search, that is, the agent $a$ chooses an action uniformly randomly among all possible actions with probability $\epsilon$, and chooses the best action with respect to the current action-value function with probability $1-\epsilon$.
As before, we consider the situation that two agents alternately learn their optimal strategies until $Q$ values converge.

We set parameters $(R,S,T,P)=(4, 0, 6, 1)$, $\eta=0.2$, and $\epsilon=0.01$.
In the numarical calculation of $Q$, we take the statistical average over $10^3$ realizations.
The initial condition of $Q$ is $Q\left( \sigma^{(1)}, \bm{\sigma}^{(0)} \right)=0$ for all $\sigma^{(1)}$ and $\bm{\sigma}^{(0)}$.

\begin{figure}[tbp]
\includegraphics[clip, width=12.0cm]{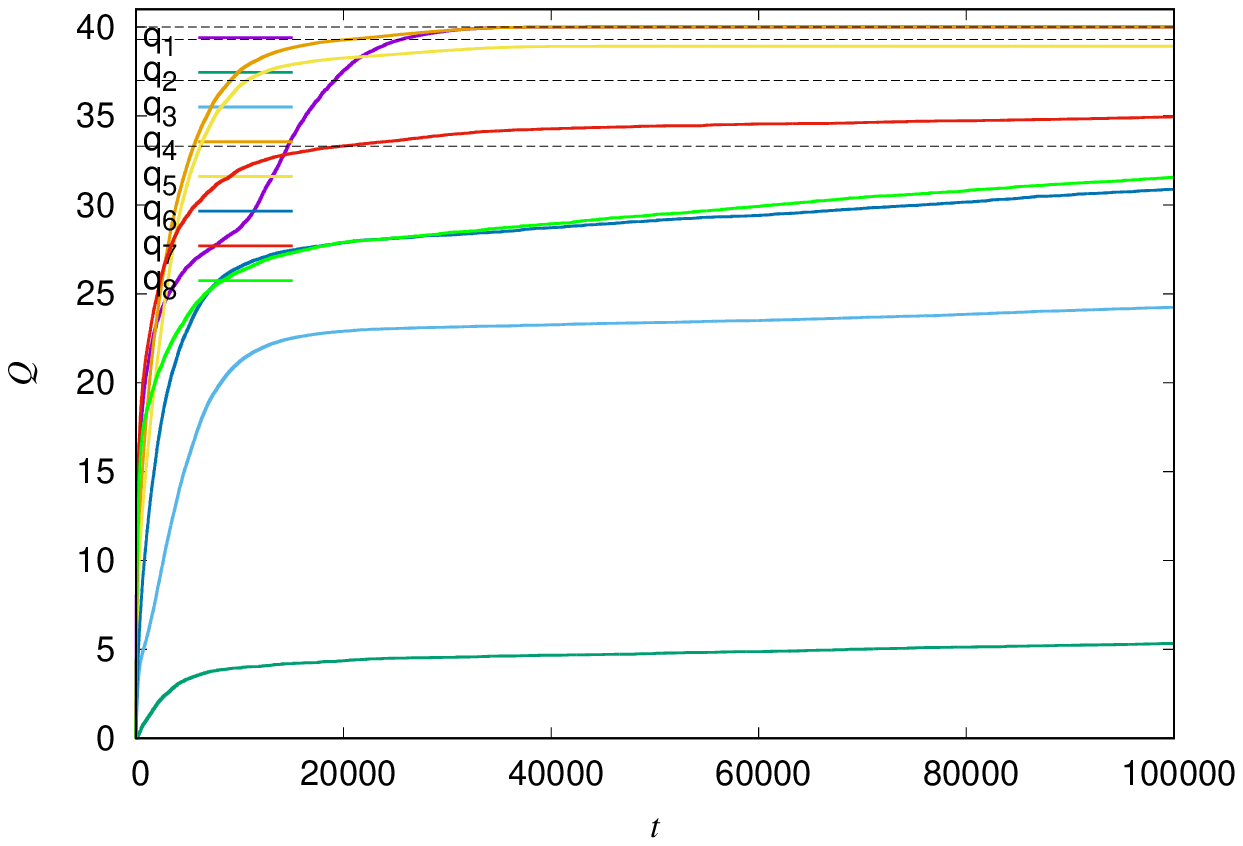}
\includegraphics[clip, width=12.0cm]{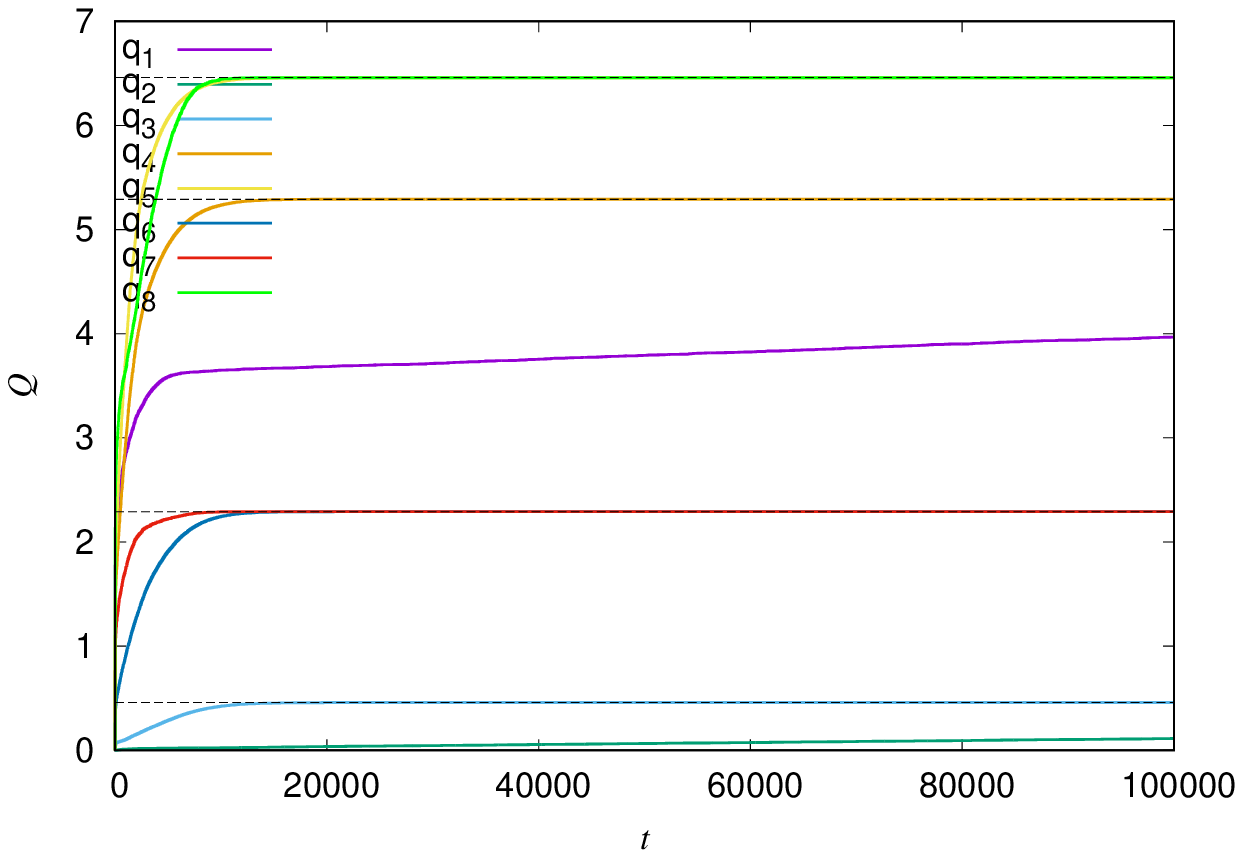}
\caption{The time evolution of $Q_1$ when the strategy of player $2$ is WSLS. (Top) The value of the discounting factor $\gamma$ is $\gamma=0.9$. The straight dash lines correspond to $40$, $39.3$, $37$, and $33.3$ from top to bottom. (Bottom) The value of the discounting factor $\gamma$ is $\gamma=0.2$. The straight dash lines correspond to $6.46$, $5.29$, $2.29$, and $0.458$ from top to bottom.}
\label{fig:wsls}
\end{figure}

In Figure \ref{fig:wsls}, we display the time evolution of $Q_1$ when the strategy of player $2$ is WSLS.
According to \ref{app:optimal}, the optimal strategy against WSLS is WSLS for $\gamma>(T-R)/(R-P)$ and All-$D$ for $\gamma<(T-R)/(R-P)$.
On the top panel of Figure \ref{fig:wsls}, we provide the numerical results for $\gamma=0.9>(T-R)/(R-P)=2/3$.
The theoretical value of $Q_1$ is also provided in \ref{app:optimal}:
\begin{eqnarray}
 q_1 = q_4 &=& \frac{1}{1-\gamma} R = 40 \\
 q_2 = q_3 &=& S + \gamma P + \frac{\gamma^2}{1-\gamma} R = 33.3 \\
 q_5 = q_8 &=& T + \gamma P + \frac{\gamma^2}{1-\gamma} R = 39.3 \\
 q_6 = q_7 &=& P + \frac{\gamma}{1-\gamma} R = 37.
\end{eqnarray}
We can expect that the numerical results converge to the theoretical value in the limit $t\rightarrow \infty$.
We emphasize that the learned strategy by player $1$ is also WSLS, which is consistent with the result in the previous section.
We remark that, as the learning proceeds, cooperation by player $1$ after the state $(C,D)$ becomes difficult to occur, which leads to the slow convergence of $Q_1(C,C,D)$.
On the bottom panel of Figure \ref{fig:wsls}, we provide the numerical results for $\gamma=0.2<(T-R)/(R-P)=2/3$.
The theoretical value of $Q_1$ is also provided in \ref{app:optimal}:
\begin{eqnarray}
 q_1 = q_4 &=& R + \frac{\gamma}{1-\gamma^2} T + \frac{\gamma^2}{1-\gamma^2} P \simeq 5.29 \\
 q_2 = q_3 &=& S + \frac{\gamma}{1-\gamma^2} P + \frac{\gamma^2}{1-\gamma^2} T \simeq 0.458 \\
 q_5 = q_8 &=& \frac{1}{1-\gamma^2} T + \frac{\gamma}{1-\gamma^2} P \simeq 6.46 \\
 q_6 = q_7 &=& \frac{1}{1-\gamma^2} P + \frac{\gamma}{1-\gamma^2} T \simeq 2.29.
\end{eqnarray}
We can expect that the numerical results also converge to the theoretical value in the limit $t\rightarrow \infty$.
For this case, the learned strategy by player $1$ is All-$D$.
Therefore, we conclude that WSLS forms the equilibrium of mutual reinforcement learning for sufficiently large $\gamma$.

\begin{figure}[tbp]
\includegraphics[clip, width=12.0cm]{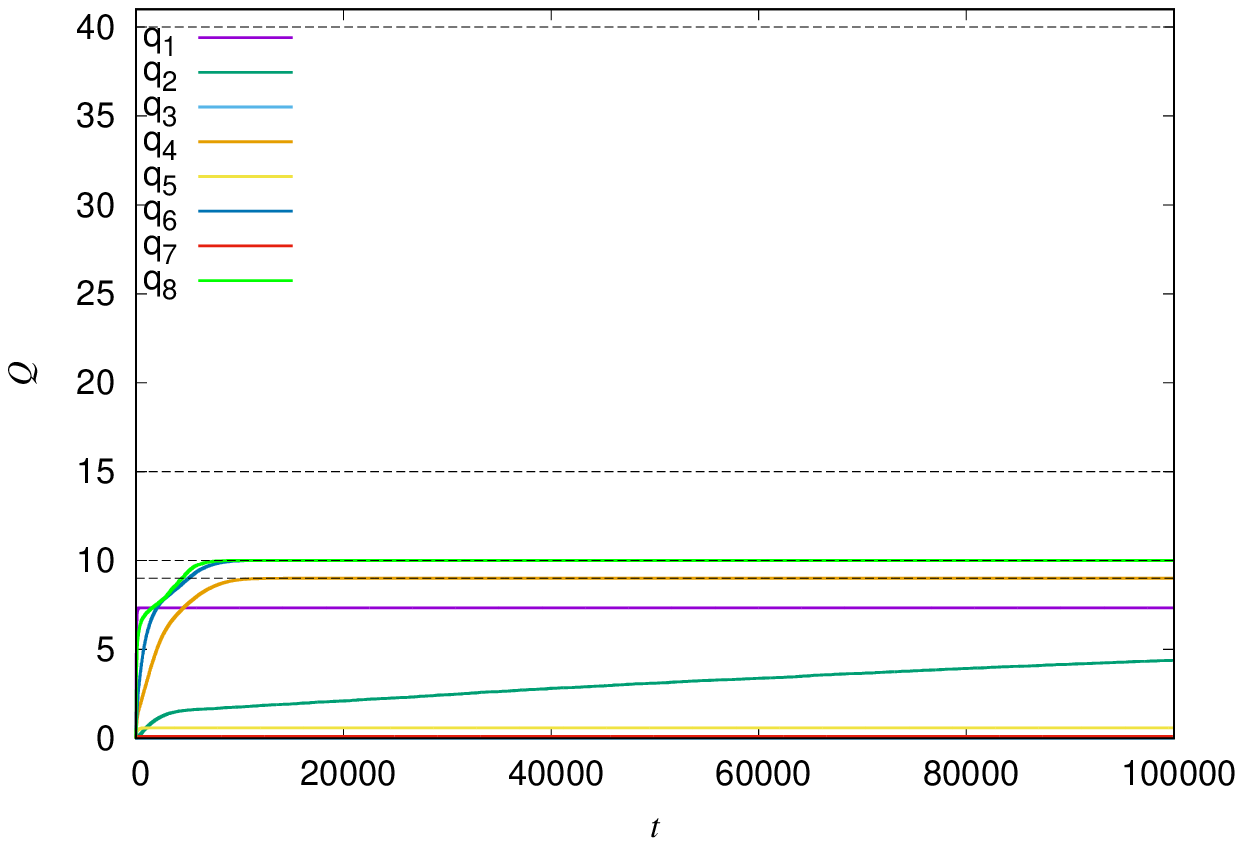}
\includegraphics[clip, width=12.0cm]{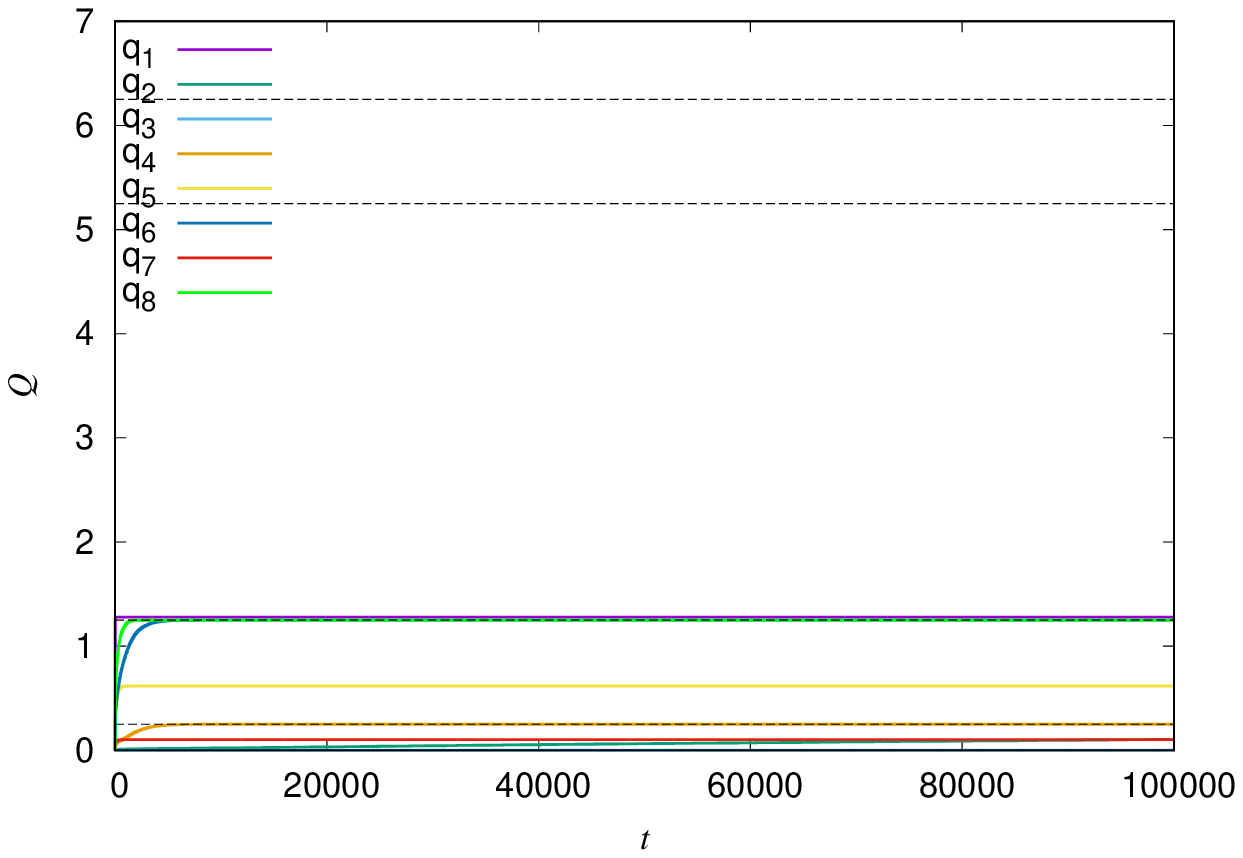}
\caption{The time evolution of $Q_1$ when the strategy of player $2$ is Grim. (Top) The value of the discounting factor $\gamma$ is $\gamma=0.9$. The straight dash lines correspond to $40$, $15$, $10$, and $9$ from top to bottom. (Bottom) The value of the discounting factor $\gamma$ is $\gamma=0.2$. The straight dash lines correspond to $6.25$, $5.25$, $1.25$, and $0.25$ from top to bottom.}
\label{fig:grim}
\end{figure}

In Figure \ref{fig:grim}, we display the time evolution of $Q_1$ when the strategy of player $2$ is Grim.
According to \ref{app:optimal}, the optimal strategy against Grim is Grim for $\gamma>(T-R)/(T-P)$ and All-$D$ for $\gamma<(T-R)/(T-P)$.
On the top panel of Figure \ref{fig:grim}, we provide the numerical results for $\gamma=0.9>(T-R)/(T-P)=2/5$.
The theoretical value of $Q_1$ is also provided in \ref{app:optimal}:
\begin{eqnarray}
 q_1 &=& \frac{1}{1-\gamma} R = 40 \\
 q_2 = q_3 = q_4 &=& S + \frac{\gamma}{1-\gamma} P = 9 \\
 q_5 &=& T + \frac{\gamma}{1-\gamma} P = 15 \\
 q_6 = q_7 = q_8 &=& \frac{1}{1-\gamma} P = 10.
\end{eqnarray}
We find that, although the learned strategy by player $1$ is Grim, there are discrepancies between the theoretical values and the numerical results for $Q_1(C,C,C)$, $Q_1(C,D,C)$, $Q_1(D,C,C)$, and $Q_1(D,D,C)$.
This is due to the property of the Grim strategy.
In our simulation, player $1$ (a learning agent against Grim) stochastically chooses $C$ or $D$.
However, once player $1$ chooses $D$, player $2$ (the agent with the Grim strategy) switches to a defector who always defects.
Therefore, the state $(D,C)$ occurs only once.
Similarly, the state $(C,C)$ occurs only while player $1$ keeps cooperating.
Therefore, the number of times that the states $(C,C)$ and $(D,C)$ occur in one trial of the infinitely repeated game cannot be large enough for the $Q$ values to converge to the theoretical values.
(It has been known that the action-value function in Q-learning converges to the true value if all state-action pairs are visited an infinite number of times \cite{SutBar2018}.)
In addition, as the learning proceeds, cooperation by player $1$ after the state $(C,D)$ becomes difficult to occur, which leads to the slow convergence of $Q_1(C,C,D)$.
On the bottom panel of Figure \ref{fig:grim}, we provide the numerical results for $\gamma=0.2<(T-R)/(T-P)=2/5$.
The theoretical value of $Q_1$ is also provided in \ref{app:optimal}:
\begin{eqnarray}
 q_1 &=& R + \gamma T + \frac{\gamma^2}{1-\gamma} P = 5.25 \\
 q_2 = q_3 = q_4 &=& S + \frac{\gamma}{1-\gamma} P = 0.25 \\
 q_5 &=& T + \frac{\gamma}{1-\gamma} P = 6.25 \\
 q_6 = q_7 = q_8 &=& \frac{1}{1-\gamma} P = 1.25.
\end{eqnarray}
We find that the learned strategy by player $1$ is Grim, although the theoretical prediction is All-$D$.
Due to the same reason as above, there are discrepancies between the theoretical values and the numerical results for $Q_1(C,C,C)$, $Q_1(C,D,C)$, $Q_1(D,C,C)$, and $Q_1(D,D,C)$.
In particular, although $Q_1(C,C,C)$ is updated as long as player $1$ keeps cooperating, $Q_1(D,C,C)$ is updated only once, that is, when player $1$ first defects.
This fact leads to the discrepancy between the theoretical prediction $Q_1(C,C,C)<Q_1(D,C,C)$ and the numerical result $Q_1(C,C,C)>Q_1(D,C,C)$.
(In order to check this conjecture, we also provide numerical results about the situation where implementation error exists in the action of player $2$, in \ref{app:error}.
These results are consistent with our conjecture.)
In addition, due to the same reason as above, the convergence of $Q_1(C,C,D)$ is slow.
Besides these facts, our numerical results are consistent with the theoretical prediction, and we conclude that Grim can form the equilibrium of mutual reinforcement learning.

\section{Conclusion}
\label{sec:conclusion}
In this paper, we theoretically investigated the situation where both players alternately use reinforcement learning to obtain their optimal memory-one strategies in the repeated prisoner's dilemma game.
We derived the symmetric solutions of the Bellman optimality equations.
We found that WSLS, the Grim strategy, and the All-$D$ strategy can form equilibrium of the mutual reinforcement learning process amongst sixteen deterministic memory-one strategies.
We checked this result by numerical simulation using $Q$-learning.
The following problems should be studied in future: (i) Whether asymmetric equilibrium points exist or not, (ii) analysis on non-deterministic strategies, and (iii) extension of our analysis to memory-two strategies.
Furthermore, extension of our analysis to the situations where the inequalities $T>R>P>S$ \cite{TanSag2007,WKJT2015,ItoTan2018,AKJIT2020} or $2R>T+S$ \cite{TanSag2007b,WakTan2011} do not hold is also a subject of future work.
In addition, elucidating the relation between equilibrium in the mutual reinforcement learning and equilibrium in evolutionary game theory \cite{SmiPri1973} is a significant problem.

\section*{Acknowledgement}
This study was supported by JSPS KAKENHI Grant Number JP20K19884.

\appendix
\section{Optimal strategy against fixed strategies}
\label{app:optimal}
In this appendix, we provide theoretical results on the deterministic optimal strategy of a learning agent against the other agent with a fixed strategy.
We regard agent $1$ and $2$ as a learning agent and an agent with a fixed strategy, respectively.
The Bellman optimality equation of the agent $1$ is Eq. (\ref{eq:optimal_1}) as before.
We consider the situation where the agent $2$ chooses the TFT strategy, the WSLS strategy, and the Grim strategy.
We introduce the notation (\ref{eq:def_q}) as before.

\subsection{Optimal strategy against TFT}
Here we consider the situation that the strategy of the agent $2$ is TFT:
\begin{eqnarray}
 \bm{T}_2 (C) &=&  \left(
\begin{array}{c}
1 \\
1 \\
0 \\
0
\end{array}
\right).
\end{eqnarray}
Then, the solution of Eq. (\ref{eq:optimal_1}) is as follows.

\subsubsection{The case $T+S<R+P$ and $\gamma>\frac{P-S}{R-S}$}
For the case, the solution is
\begin{eqnarray}
 q_1 = q_2 &=& \frac{1}{1-\gamma} R \\
 q_3 = q_4 &=& S + \frac{\gamma}{1-\gamma} R \\
 q_5 = q_6 &=& T + \gamma S + \frac{\gamma^2}{1-\gamma} R \\
 q_7 = q_8 &=& P + \gamma S + \frac{\gamma^2}{1-\gamma} R
\end{eqnarray}
and because $q_1>q_5$, $q_2>q_6$, $q_3>q_7$, and $q_4>q_8$, the optimal strategy is All-$C$.

\subsubsection{The case $T+S<R+P$ and $\frac{T-R}{T-P}<\gamma<\frac{P-S}{R-S}$}
For the case, the solution is
\begin{eqnarray}
 q_1 = q_2 &=& \frac{1}{1-\gamma} R \\
 q_3 = q_4 &=& S + \frac{\gamma}{1-\gamma} R \\
 q_5 = q_6 &=& T + \frac{\gamma}{1-\gamma} P \\
 q_7 = q_8 &=& \frac{1}{1-\gamma} P
\end{eqnarray}
and because $q_1>q_5$, $q_2>q_6$, $q_3<q_7$, and $q_4<q_8$, the optimal strategy is Repeat.

\subsubsection{The case $T+S<R+P$ and $\gamma<\frac{T-R}{T-P}$}
For the case, the solution is
\begin{eqnarray}
 q_1 = q_2 &=& R + \gamma T + \frac{\gamma^2}{1-\gamma} P \\
 q_3 = q_4 &=& S + \gamma T + \frac{\gamma^2}{1-\gamma} P \\
 q_5 = q_6 &=& T + \frac{\gamma}{1-\gamma} P \\
 q_7 = q_8 &=& \frac{1}{1-\gamma} P
\end{eqnarray}
and because $q_1<q_5$, $q_2<q_6$, $q_3<q_7$, and $q_4<q_8$, the optimal strategy is All-$D$.

\subsubsection{The case $T+S>R+P$ and $\gamma>\frac{T-R}{R-S}$}
For the case, the solution is
\begin{eqnarray}
 q_1 = q_2 &=& \frac{1}{1-\gamma} R \\
 q_3 = q_4 &=& S + \frac{\gamma}{1-\gamma} R \\
 q_5 = q_6 &=& T + \gamma S + \frac{\gamma^2}{1-\gamma} R \\
 q_7 = q_8 &=& P + \gamma S + \frac{\gamma^2}{1-\gamma} R
\end{eqnarray}
and because $q_1>q_5$, $q_2>q_6$, $q_3>q_7$, and $q_4>q_8$, the optimal strategy is All-$C$.

\subsubsection{The case $T+S>R+P$ and $\frac{P-S}{T-P}<\gamma<\frac{T-R}{R-S}$}
For the case, the solution is
\begin{eqnarray}
 q_1 = q_2 &=& R + \frac{\gamma}{1-\gamma^2}T + \frac{\gamma^2}{1-\gamma^2} S \\
 q_3 = q_4 &=& \frac{1}{1-\gamma^2}S + \frac{\gamma}{1-\gamma^2}T \\
 q_5 = q_6 &=& \frac{1}{1-\gamma^2}T + \frac{\gamma}{1-\gamma^2}S \\
 q_7 = q_8 &=& P + \frac{\gamma}{1-\gamma^2}S + \frac{\gamma^2}{1-\gamma^2} T
\end{eqnarray}
and because $q_1<q_5$, $q_2<q_6$, $q_3>q_7$, and $q_4>q_8$, the optimal strategy is anti-Repeat.

\subsubsection{The case $T+S>R+P$ and $\gamma<\frac{P-S}{T-P}$}
For the case, the solution is
\begin{eqnarray}
 q_1 = q_2 &=& R + \gamma T + \frac{\gamma^2}{1-\gamma} P \\
 q_3 = q_4 &=& S + \gamma T + \frac{\gamma^2}{1-\gamma} P \\
 q_5 = q_6 &=& T + \frac{\gamma}{1-\gamma} P \\
 q_7 = q_8 &=& \frac{1}{1-\gamma} P
\end{eqnarray}
and because $q_1<q_5$, $q_2<q_6$, $q_3<q_7$, and $q_4<q_8$, the optimal strategy is All-$D$.

\subsection{Optimal strategy against WSLS}
Here we consider the situation that the strategy of the agent $2$ is WSLS:
\begin{eqnarray}
 \bm{T}_2 (C) &=&  \left(
\begin{array}{c}
1 \\
0 \\
0 \\
1
\end{array}
\right).
\end{eqnarray}
Then, the solution of Eq. (\ref{eq:optimal_1}) is as follows.

\subsubsection{The case $T+P<2R$ and $\gamma>\frac{T-R}{R-P}$}
For the case, the solution is
\begin{eqnarray}
 q_1 = q_4 &=& \frac{1}{1-\gamma} R \\
 q_2 = q_3 &=& S + \gamma P + \frac{\gamma^2}{1-\gamma} R \\
 q_5 = q_8 &=& T + \gamma P + \frac{\gamma^2}{1-\gamma} R \\
 q_6 = q_7 &=& P + \frac{\gamma}{1-\gamma} R
\end{eqnarray}
and because $q_1>q_5$, $q_2<q_6$, $q_3<q_7$, and $q_4>q_8$, the optimal strategy is WSLS.

\subsubsection{The case $T+P<2R$ and $\gamma<\frac{T-R}{R-P}$}
For the case, the solution is
\begin{eqnarray}
 q_1 = q_4 &=& R + \frac{\gamma}{1-\gamma^2} T + \frac{\gamma^2}{1-\gamma^2} P \\
 q_2 = q_3 &=& S + \frac{\gamma}{1-\gamma^2} P + \frac{\gamma^2}{1-\gamma^2} T \\
 q_5 = q_8 &=& \frac{1}{1-\gamma^2} T + \frac{\gamma}{1-\gamma^2} P \\
 q_6 = q_7 &=& \frac{1}{1-\gamma^2} P + \frac{\gamma}{1-\gamma^2} T
\end{eqnarray}
and because $q_1<q_5$, $q_2<q_6$, $q_3<q_7$, and $q_4<q_8$, the optimal strategy is All-$D$.

\subsubsection{The case $T+P>2R$}
For the case, the solution is
\begin{eqnarray}
 q_1 = q_4 &=& R + \frac{\gamma}{1-\gamma^2} T + \frac{\gamma^2}{1-\gamma^2} P \\
 q_2 = q_3 &=& S + \frac{\gamma}{1-\gamma^2} P + \frac{\gamma^2}{1-\gamma^2} T \\
 q_5 = q_8 &=& \frac{1}{1-\gamma^2} T + \frac{\gamma}{1-\gamma^2} P \\
 q_6 = q_7 &=& \frac{1}{1-\gamma^2} P + \frac{\gamma}{1-\gamma^2} T
\end{eqnarray}
and because $q_1<q_5$, $q_2<q_6$, $q_3<q_7$, and $q_4<q_8$, the optimal strategy is All-$D$.

\subsection{Optimal strategy against Grim}
Here we consider the situation that the strategy of the agent $2$ is Grim:
\begin{eqnarray}
 \bm{T}_2 (C) &=&  \left(
\begin{array}{c}
1 \\
0 \\
0 \\
0
\end{array}
\right).
\end{eqnarray}
Then, the solution of Eq. (\ref{eq:optimal_1}) is as follows.

\subsubsection{The case $\gamma>\frac{T-R}{T-P}$}
For the case, the solution is
\begin{eqnarray}
 q_1 &=& \frac{1}{1-\gamma} R \\
 q_2 = q_3 = q_4 &=& S + \frac{\gamma}{1-\gamma} P \\
 q_5 &=& T + \frac{\gamma}{1-\gamma} P \\
 q_6 = q_7 = q_8 &=& \frac{1}{1-\gamma} P
\end{eqnarray}
and because $q_1>q_5$, $q_2<q_6$, $q_3<q_7$, and $q_4<q_8$, the optimal strategy is Grim.

\subsubsection{The case $\gamma<\frac{T-R}{T-P}$}
For the case, the solution is
\begin{eqnarray}
 q_1 &=& R + \gamma T + \frac{\gamma^2}{1-\gamma} P \\
 q_2 = q_3 = q_4 &=& S + \frac{\gamma}{1-\gamma} P \\
 q_5 &=& T + \frac{\gamma}{1-\gamma} P \\
 q_6 = q_7 = q_8 &=& \frac{1}{1-\gamma} P
\end{eqnarray}
and because $q_1<q_5$, $q_2<q_6$, $q_3<q_7$, and $q_4<q_8$, the optimal strategy is All-$D$.

\section{Numerical results under implementation error}
\label{app:error}
In this appendix, we provide numerical results about the situation where implementation error exists in the action of a player.
The setup of numerical simulation is the same as one in Section \ref{sec:numerical}.
We assume that the strategy of player $2$ is Grim with implementation error, which takes wrong action with small probability $10^{-2}$.
The strategy of player $1$ is learned by Q-learning.

\begin{figure}[tbp]
\includegraphics[clip, width=12.0cm]{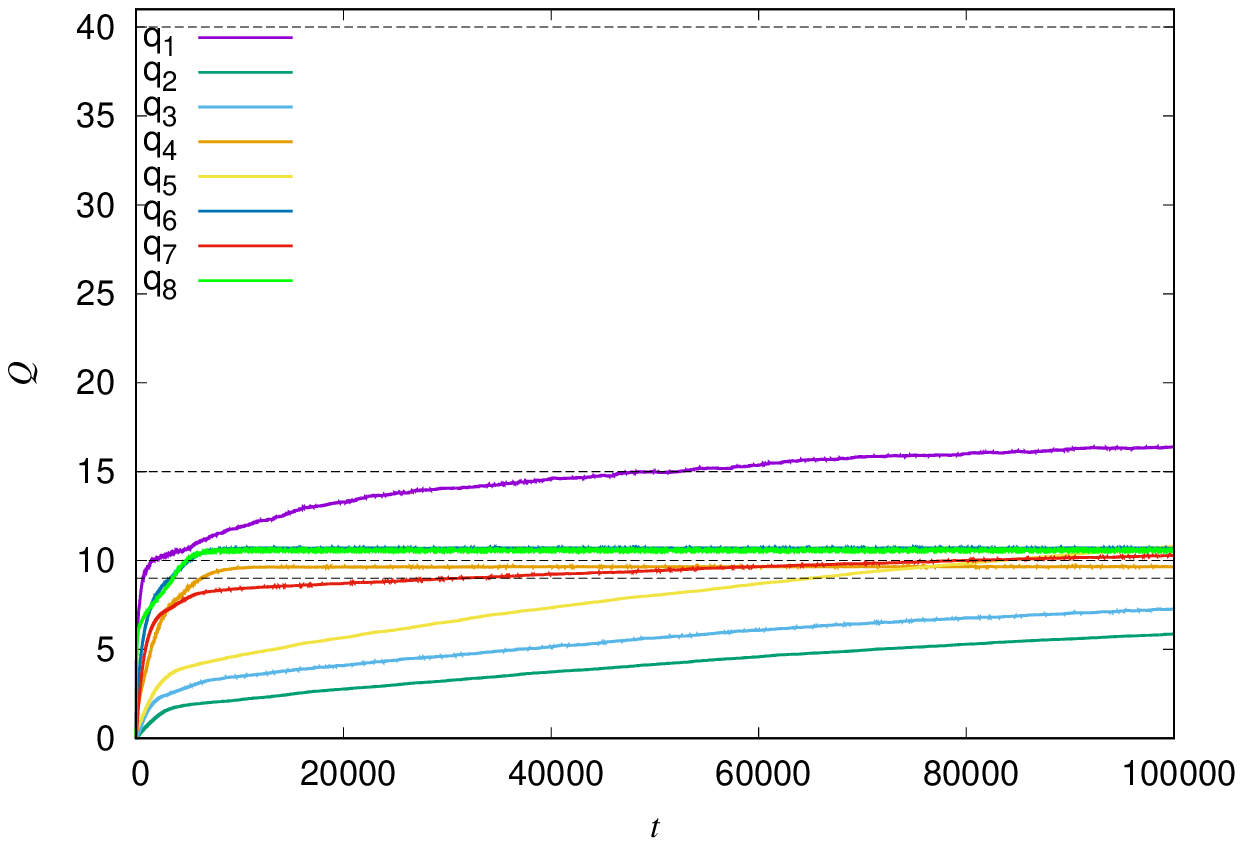}
\includegraphics[clip, width=12.0cm]{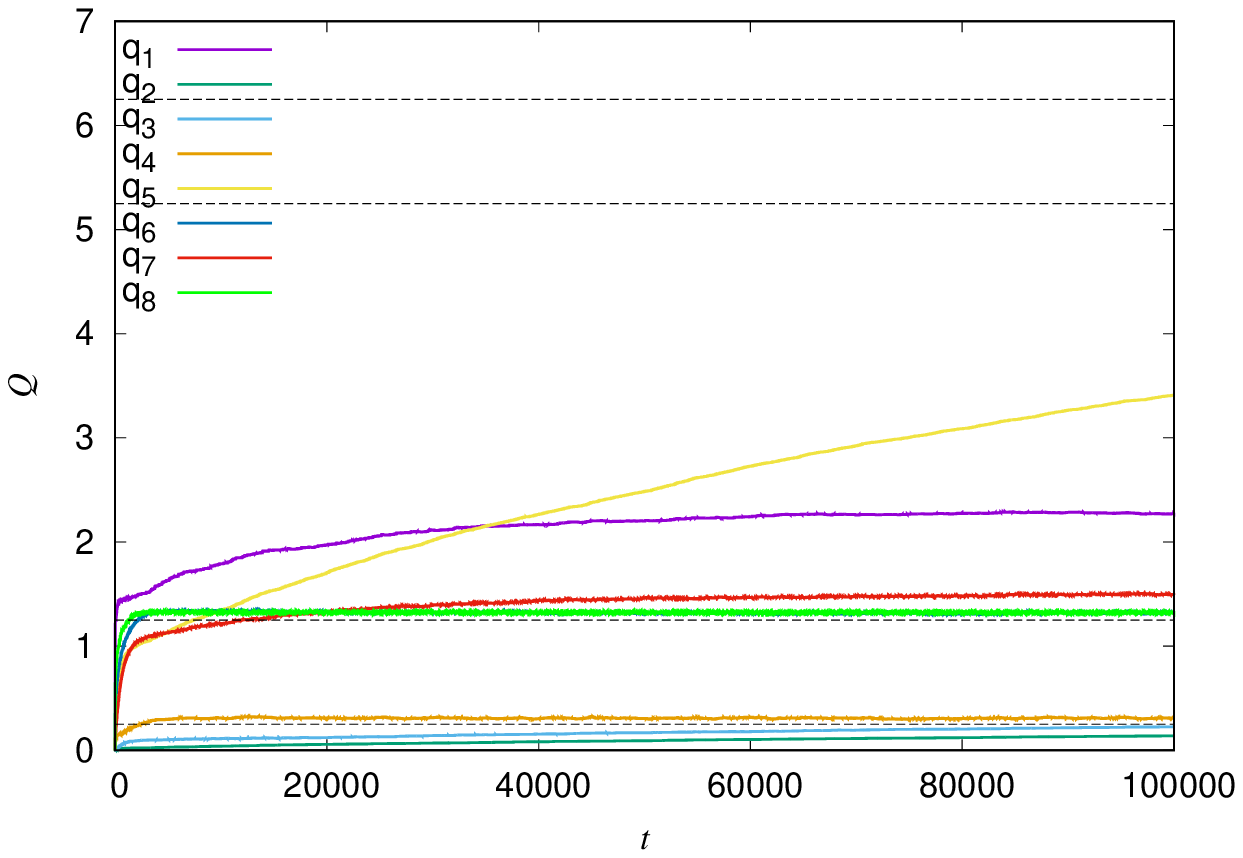}
\caption{The time evolution of $Q_1$ when the strategy of player $2$ is Grim with implementation error. (Top) The value of the discounting factor $\gamma$ is $\gamma=0.9$. The straight dash lines correspond to $40$, $15$, $10$, and $9$ from top to bottom. (Bottom) The value of the discounting factor $\gamma$ is $\gamma=0.2$. The straight dash lines correspond to $6.25$, $5.25$, $1.25$, and $0.25$ from top to bottom.}
\label{fig:grim_error}
\end{figure}

In Figure \ref{fig:grim_error}, we display the time evolution of $Q_1$ when the strategy of player $2$ is Grim with implementation error.
We can see that the learned strategy of player $1$ is Grim for $\gamma=0.9$, and All-$D$ for $\gamma=0.2$, in contrast to the case without implementation error in Figure \ref{fig:grim}, where the learned strategy is Grim for both $\gamma=0.9$ and $\gamma=0.2$.
This is because Grim with implementation error is not irreversible, although Grim is irreversible, and $Q_1(C,C,C)$ and $Q_1(D,C,C)$ are updated sufficiently many times.

\section*{References}

\bibliography{RL_RPD}

\end{document}